\documentstyle[prl,multicol,aps,epsfig]{revtex}
\begin {document}
\title{Qubit measurements with a double-dot detector}
\author{T. Gilad and S.A. Gurvitz}
\address{Department of Particle Physics, Weizmann Institute of Science, Rehovot 76100,
Israel}
\date{\today}
\vspace{2cm} \maketitle
\begin{abstract}
We propose to monitor a qubit with a double-dot (DD)
resonant-tunneling detector, which can operate at higher
temperatures than a single-dot detector. In order to assess the
effectiveness of this device, we derive rate equations for the
density matrix of the entire system. We show that the
signal-to-noise ratio can be greatly improved by a proper choice of
the parameters and location of the detector. We demonstrate that
quantum interference effects within the DD detector play an
important role in the measurement. Surprisingly, these effects
produce a systematic measurement error, even when the entire system
is in a stationary state.
\end{abstract}
\maketitle

\hspace{1.5 cm} PACS:  73.50.-h, 73.23.-b, 03.65.Yz.
\begin{multicols}{2}
The single electron transistor (SET) is a sensitive device for
quantum measurements\cite{dev,moz1,lu}. It can be used as a monitor
of a charge qubit, provided that the energy level $E_0$ carrying the
current is close to the Fermi levels of the reservoirs $\mu_{L,R}$
[see Fig.~1(A,A')]. Then due to the electrostatic repulsion $U$
between the electrons, the SET current drops when the qubit is in
the state $E_2$, as in Fig.~1(A').

It is clear that one needs very low reservoir temperatures in order
to use the SET as a sensitive detector. This requirement, however,
can be weakened by taking a double-dot (DD) for monitoring the qubit
state, Fig.~1(B,B'). Similar to the SET, the DD current decreases
sharply whenever the electron of the qubit is close to one of the
dots, Fig.~1(B').  In contrast with the SET, however, the
reservoirs' temperature ($T$) will not affect the current if
$\mu_L-T\gg E_0\gg\mu_R+T$\cite{spr,fn0}.
\begin{figure}
{\centering{\epsfig{figure=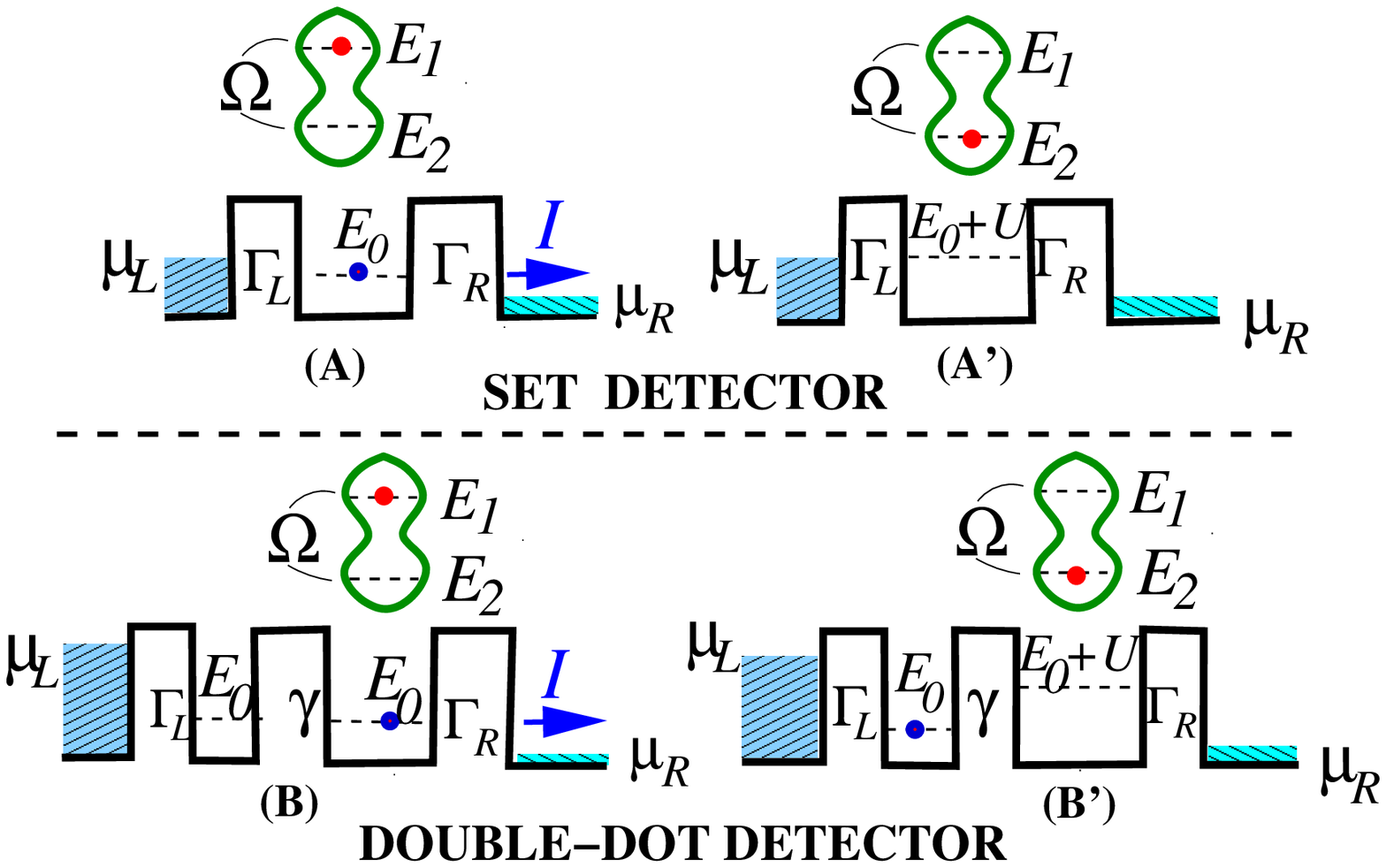,height=6cm,width=8cm,angle=0}}}
{\bf Fig.~1:} The qubit measurements using the SET detector (A,A')
and the DD detector (B,B'). The detector current $I$ drops down
whenever the state $E_2$ of the qubit is occupied. $\Gamma_{L,R}$
represent tunneling rates to the reservoirs and $\gamma$ denotes the
interdot coupling.
\end{figure}

Because of quantum interference effects, the dynamics of the
measurement process using the DD detector is more complicated than
with the SET detector. An electron flowing through the DD can be
trapped in a linear superposition of the dot states. As a result,
quantum interference could modify the signal in such a way that 
the DD cannot monitor the qubit. It is necessary to
analyze the influence of the DD  on the qubit motion (and vice
versa) in order to establish the optimal conditions for utilizing
the DD as an effective quantum detector. This can be done by solving
the Schr\"{o}dinger equation describing the combined system of qubit
and detector.

In fact, the setup shown in Fig.~1(B,B') represents a generic class
of non-demolition quantum measurements where a measured system 
interacts with only one state of the apparatus, while the apparatus 
may be in a superposition of states. This can take place in many devices 
based on interference, for instance in Electronic Mach-Zehnder 
Interferometer\cite{Ji,fn01}.  
We therefore believe that our analysis of a qubit
interacting with the DD detector can be useful for understanding
many different quantum measurements.

Let us describe the entire setup shown in Fig.~1(B,B') by the
 tunneling Hamiltonian $H=H_q+H_{dd}+H_{int}$, where
\begin{eqnarray}
&&H_{q}=E_1a_1^\dagger a_1+E_2a_2^\dagger a_2+\Omega(a_1^\dagger a_2
+a_2^\dagger a_1)\, ,\nonumber\\
&&H_{dd}=H_0+E_0(c_1^\dagger c_1+c_2^\dagger c_2)+\gamma(c_1^\dagger
c_2
+c_2^\dagger c_1)\nonumber\\
&&+\sum_\lambda(\Omega_\lambda^Lc^\dagger_1c_\lambda^L
+\Omega_\lambda^Rc^\dagger_2 c_\lambda^R+H.c.)
+\bar U_{12}c_1^\dagger c_1c_2^\dagger c_2\nonumber\\
&&H_{int}=Ua_2^\dagger a_2c_2^\dagger c_2 \label{a1}
\end{eqnarray}
are the qubit and the DD Hamiltonians, and $H_{int}$ is their
interaction. Here $a^\dagger(a)$ is the creation (annihilation)
operator for the electron in the qubit and $c^\dagger(c)$ is the
same operator for the DD; $\Omega$ is the coupling between the
states $|a^\dagger_{1,2}|0\rangle$ of the qubit, and $\gamma$ is the
coupling between the states $|c^\dagger_{1,2}|0\rangle$ of the DD.
The Hamiltonian $H_0=\sum_\lambda [E_\lambda^L(c_\lambda^L)^\dagger
c_\lambda^L +E_\lambda^R (c_\lambda^R)^\dagger c_\lambda^R]$
describes the reservoirs, where $\Omega_\lambda^{L,R}$ are the
couplings between the right and left dots with the right and left
reservoirs. We assume weak energy dependence of these couplings,
$\Omega_\lambda^{L,R}\simeq\Omega_{L,R}$. Then the corresponding
tunneling rates are $\Gamma_{L,R}=2\pi\rho_{L,R}\Omega_{L,R}^2$,
where $\rho_{L,R}$ are the density of states in the reservoirs. The
latter quantities are also weakly dependent of energy. The last term
in $H_{dd}$ describes the inter-dot repulsion. For simplicity we
consider electrons as spinless fermions.

Using the technique developed in Ref.\cite{g,g2} for the case of
large bias voltage, $V=\mu_L-\mu_R$, we can partially trace out the
reservoir states in the equation of motion for the density matrix of
an entire system, $i\dot\varrho =[H,\varrho]$. As a result we arrive
at the Bloch-type rate equations for the reduced density-matrix,
$\sigma_{ij}^{n}(t)$, describing the qubit-detector evolution, where
the indices $i,j$ denote all available discrete states of the
detector-qubit system and $n$ denotes the number of electrons which
have arrived at the right reservoir by time $t$. In our case,
Fig.~1(B,B'), the available discrete states are labeled ($a,b,c,d$),
denoting the cases that the DD is empty ($a$), the left dot of the
DD system is occupied ($b$), the right dot of the DD system is
occupied ($c$), and both dots are occupied ($d$), while the electron
of the qubit occupies the level $E_1$ (see Fig.~1B).
Correspondingly, ($a',b',c',d'$) denote the same states but where
the electron of the qubit occupies the level $E_2$ (see Fig.~1B').
If the inter-dot repulsion is large, $\bar U_{12}\gg V$, the states
$d,d'$ do not contribute terms to the equations of motion. We obtain
in this case\cite{g,g2,gb}:
\begin{mathletters}
\label{a2}
\begin{eqnarray}
\label{a2a} &&\dot\sigma_{aa}^n=-\Gamma_L\,\sigma_{aa}^n+\Gamma_R\,
\sigma _{cc}^{n-1} +i\Omega (\sigma_{aa'}^n
-\sigma_{a'a}^n)\\
\label{a2b}&&\dot\sigma_{a'a'}^n =-\Gamma_L\, \sigma_{a'a'}^n
+\Gamma_R\,
\sigma _{c'c'}^{n-1} +i\Omega (\sigma _{a'a}^n -\sigma _{aa'}^n )\\
\label{a2c} &&\dot\sigma_{bb}^n =\Gamma_L\, \sigma _{aa}^n +i\Omega
(\sigma _{bb'}^n -\sigma _{b'b}^n )+i\gamma (\sigma _{bc}^n -\sigma
_{cb}^n )\\ \label{a2d} &&\dot\sigma _{b'b'}^n =\Gamma _L\, \sigma
_{a'a'}^n +i\Omega (\sigma _{b'b}^n -\sigma _{bb'}^n )+i\gamma
(\sigma_{b'c'}^n -\sigma _{c'b'}^n )\\
\label{a2e} &&\dot\sigma _{cc}^n =-\Gamma _R\, \sigma _{cc}^n
+i\Omega (\sigma _{cc'}^n -\sigma _{c'c}^n) +i\gamma (\sigma _{cb}^n
-\sigma _{bc}^n )\\
\label{a2f} &&\dot\sigma _{c'c'}^n =-\Gamma _R \sigma _{c'c'}^n
+i\Omega (\sigma _{c'c}^n -\sigma _{cc'}^n )+i\gamma (\sigma
_{c'b'}^n
-\sigma _{b'c'}^n )\\
\label{a2g} &&\dot\sigma _{aa'}^n =i\Omega (\sigma _{aa}^n -\sigma
_{a'a'}^n )-\Gamma _L\, \sigma _{aa'}^n +\Gamma _R\, \sigma _{cc'}^{n-1}\\
\label{a2h} &&\dot\sigma _{bb'}^n =i\Omega (\sigma _{bb}^n -\sigma
_{b'b'}^n )+i\gamma (\sigma _{bc'}^n -\sigma _{cb'}^n )+\Gamma _L\,
\sigma
_{aa'}^n\\
\label{a2i} &&\dot\sigma _{bc}^n =i\Omega (\sigma _{bc'}^n -\sigma
_{b'c}^n )+i\gamma (\sigma _{bb}^n -\sigma _{cc}^n )-\frac{\Gamma _R
}{2}\sigma _{bc}^n\\ \nonumber
 &&\dot\sigma _{b'c'}^n =iU\sigma _{b'c'}^n +i\Omega (\sigma
_{b'c}^n -\sigma _{bc'}^n )+i\gamma (\sigma _{b'b'}^n -\sigma
_{c'c'}^n )\\ \label{a2j}
&&~~~~~~~~~~~~~~~~~~~~~~~~~~~~~~~~~~~~~~~~~~~~~~~~
-\frac{\Gamma _R }{2}\sigma _{b'c'}^n\\
\nonumber&&\dot\sigma _{cc'}^n =iU\sigma _{cc'}^n +i\Omega (\sigma
_{cc}^n -\sigma _{c'c'}^n )+i\gamma (\sigma _{cb'}^n -\sigma
_{bc'}^n )\\&&~~~~~~~~~~~~~~~~~~~~~~~~~~~~~~~~~~~~~~~~~~~~~
-\Gamma_R \sigma _{cc'}^n\\ \nonumber &&\dot\sigma _{bc'}^n =
iU\sigma _{bc'}^n +i\Omega (\sigma _{bc}^n
-\sigma _{b'c'}^n )+i\gamma (\sigma _{bb'}^n -\sigma _{cc'}^n )\\
\label{a2l} &&~~~~~~~~~~~~~~~~~~~~~~~~~~~~~~~~~~~~~~~~~~~~~~
 -\frac{\Gamma _R }{2}\sigma _{bc'}^n\\ \label{a2m}
&&\dot\sigma _{cb'}^n =i\Omega (\sigma _{cb}^n -\sigma _{c'b'}^n
)+i\gamma (\sigma _{cc'}^n -\sigma _{bb'}^n ) -\frac{\Gamma_R
}{2}\sigma _{cb'}^n
\end{eqnarray}
\end{mathletters}
Note that these equations are obtained from the original many-body
equations $i\dot\varrho =[H,\varrho]$ without the explicit use of
any Markov-type or weak-coupling approximations in the case of large
bias voltage, $V\gg\Gamma_{L,R},\, U$\cite{g,g2}. There are no other
limitation on $U$, in contrast with our analysis of the SET
detector\cite{gb}. 

Equations~(\ref{a2}) are different from the standard master equations, 
describing quantum system interacting with the environment (detector) 
by keeping track of the environment variables. In our case this is 
the number of electrons
$(n)$ arriving the collector. This allows us to find the time evolution of 
the qubit and the detector at once. For
instance, the qubit behavior is described by the (reduced) density
matrix $\sigma_{q}(t)\equiv \{\sigma_{\alpha\beta}(t)\}$ with
$\alpha,\beta=\{1,2\}$, where $\sigma_{11}=\sum_n
(\sigma_{aa}^n+\sigma_{bb}^n+\sigma_{cc}^n)$, $\sigma_{12}=\sum_n
(\sigma_{aa'}^n+\sigma_{bb'}^n+\sigma_{cc'}^n)$ and
$\sigma_{22}=1-\sigma_{11}$.

On the other hand, by tracing out the qubit variables we obtain the
probability of finding $n$ electrons which have arrived at the
collector, $P_n(t)=\sum_j\sigma_{jj}^{n}(t)$.  This quantity allows
us to determine the average detector current and its shot-noise
spectrum. The former is given by
\begin{equation}
I(t)=e\sum_n n\dot P_n(t)=e\Gamma_R\,\sigma_R(t), \label{a3}
\end{equation}
where $\sigma_R(t)=\sum_n[\sigma_{cc}^n(t)+\sigma_{c'c'}^n(t)]$ is
the probability that the right dot is occupied. The shot-noise 
spectrum, $S(\omega )$, is obtained from the McDonald formula
\cite{mac,moz2}
\begin{equation}
S(\omega) = 2e^2\omega \int_0^\infty dt \sin (\omega t) \sum_n
n^2\dot P_n(t)\, , \label{a4}
\end{equation}
One finds from Eqs.~(\ref{a2}), (\ref{a4}) that
\begin{equation}
S(\omega) = 2e^2\omega\Gamma_R {\mbox{Im}}\,[Z_{cc}(\omega )
+Z_{c'c'}(\omega )]\, , \label{a5}
\end{equation}
where $Z_{ij}(\omega) = \int_0^\infty
\sum_n(2n+1)\sigma_{ij}^n(t)\exp (i\omega t)dt$. These quantities
are obtained directly from Eqs.~(\ref{a2}) which are reduced to a
system of linear algebraic equations after the corresponding
integration over $t$\cite{fn1}.

Consider first the static qubit, $\Omega =0$. Solving
Eqs.~(\ref{a2}) for this case one finds that the stationary current,
$\bar I=I(t\to\infty )$ obtains the value $\bar
I_1=\Gamma_R\bar\sigma_R(U=0)$ when the qubit is in the state $E_1$,
and $\bar I_2=\Gamma_R\bar\sigma_R(U)$ when the qubit is in the
state $E_2$, Fig.~1(B,B'), where\begin{equation} \label{a6}
\bar\sigma_R(U)={\gamma^2\over U^2+{\Gamma_R^2\over
4}+\gamma^2\left(2+{\Gamma_R\over\Gamma_L}\right)}\, .
\end{equation}
As expected, the detector current decreases whenever the electron of
the qubit is close to the DD detector. Consider now $\Omega\not =0$.
We assume that for an ``ideal'' detector its average current would
follow the qubit motion\cite{fn2},
\begin{equation}\label{a7}
I(t)=\bar {I}_1\sigma_{11}(t)+\bar{I}_2[1-\sigma_{11}(t)]\, .
\end{equation}
This condition, however, cannot be fully met since the detector's
response is limited by the rate of tunneling from the right dot to
the collector. Nevertheless, if this transition is fast enough
compared to the qubit frequency, $\Gamma_R\gg\Omega$, one expects to
approach Eq.~(\ref{a7}).

Let us compare $\sigma_{11}(t)$ with the average ``signal,''
$[I(t)-\bar{I}_2]/\Delta\bar I$, where $\Delta
\bar{I}=\bar{I}_1-\bar{I}_2$. The results of our calculations for
$\gamma =\Omega$ are presented in Fig.~2. The initial conditions
correspond to the qubit electron in the upper dot and the detector
current $I(t=0)=\bar{I}_1$. One finds that the detector does not
follow the qubit oscillations well when $\Gamma_R =\gamma$,
Fig.~2(a). On the other hand, in Fig.~2(b) where $\Gamma_R\gg \gamma
$, the detector performance is much improved, in accordance with our
arguments.
\begin{figure}
{\centering{\epsfig{figure=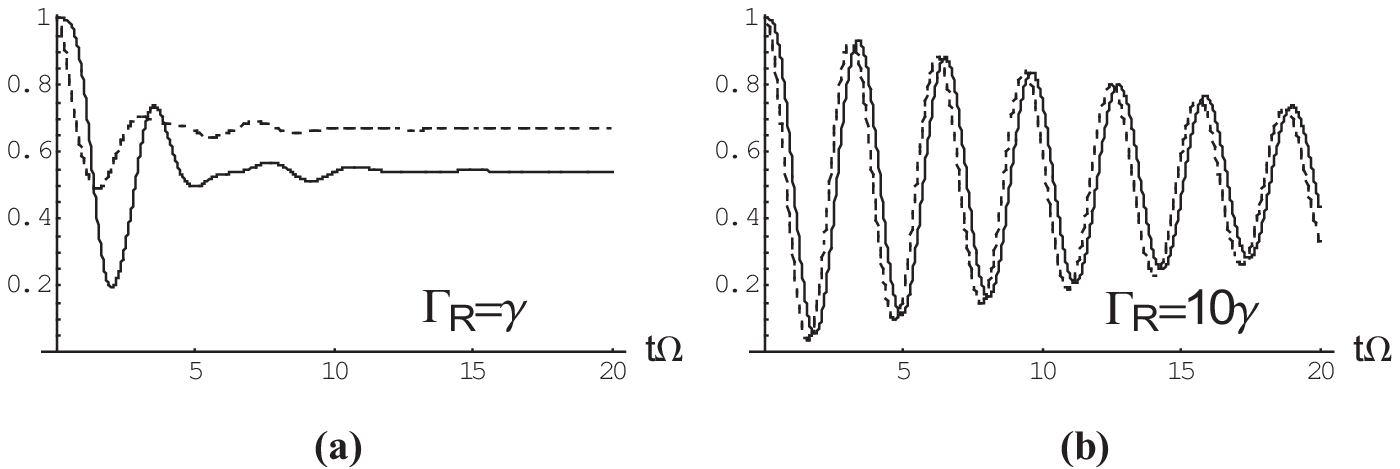,height=3cm,width=8.5cm,angle=0}}}
{\bf Fig.~2:} The probability of finding the qubit in the state
$E_1$, (dashed line) compared with the average detector signal,
$[I(t)-\bar{I}_2]/\Delta\bar I$ (solid line) as a function of time
for $\gamma =\Omega$ and  $U=5\Omega,\, \Gamma_L=5\Omega$.
\end{figure}

The results displayed in Fig.~2(a) are surprising. One expects that
in the steady-state limit ($t\to\infty$) the average detector
current should be distributed between the values $\bar {I}_1$ and
$\bar {I}_2$, with probabilities $\sigma_{11}$ and $1-\sigma_{11}$
to find the qubit in the states $E_{1,2}$, respectively.
Equation~(\ref{a7}) should thus always hold in the limit of
$t\to\infty$ for any such device (for instance, the SET detector
\cite{gb}). In the case of the DD detector, however, Eq.~(\ref{a7})
does not hold in the steady-state limit, as seen in Fig.~2(a). In
fact, this can be obtained analytically in the limit of small $U$ by
expanding the stationary current, $\bar I=I(t\to\infty )$,
Eq.~(\ref{a3}), in powers of $U$. One finds for the detector's
signal:
\begin{equation}
\label{a7a} {\bar I-\bar{I}_2\over \Delta\bar I} = \left [1+{2\over
4+(\Gamma_R /2\Omega )^2}+O(U^2)\right ]\sigma_{11}(t\to\infty )
\end{equation}
It follows from this expression that a mismatch between the signal
and the qubit ($\sigma_{11}$) survives even in the limit $U\to 0$.
In this case it depends only on the ratio $\Gamma_R /\Omega$. The
other detector parameters $\gamma$ and~$\Gamma_L$ enter only in the
term proportional to $U^2$.

So where is the ``hidden'' probability that is responsible for the
systematic error in the qubit measurements? It can be recovered in
the linear superposition of the detector and qubit states. The DD
current flows via two discrete energy levels, $E_0$ and $E_0+U$. A
carrier wave function thus proceeds through a linear superposition
of these states, $b(b')$ and $c(c')$. The qubit is itself a two
level system described by superposition. These different
superpositions involve the same states ($b,b',c,c'$) of the entire
system, and hence are entangled. This is reflected in the
off-diagonal terms $\sigma_{bc'}$ and $\sigma_{b'c}$,
Eqs.~(\ref{a2}). As a result the superposition of qubit states
(qubit's ``phase'')  directly affects the DD dynamics, leading to a
violation of Eq.~(\ref{a7}). This would happen even in the limit
$U\to\infty$. In this case the state $(c')$ disappears from
Eqs.~(\ref{a2}), but the off-diagonal term $\sigma_{cb'}$ would
still survive in the limit $t\to\infty$. One should note that in the
case of the SET such entanglement cannot occur. Therefore
Eq.~(\ref{a7}) holds for the SET, even though
$\sigma_{12}(t\to\infty)\not =0$.

We find from Fig.~2 and Eq.~(\ref{a7}) that the detector's
performance improves when $\Gamma_R\gg\Omega$. At the same time,
however, its average signal decreases. In order to assess the
detector's efficiency this signal should be compared with its noise.
An appropriate measure of the detector efficiency is the integrated
signal to noise ratio \cite{bra,moz1}, $s/n =\int_{-\infty}^\infty
[{|I_{sig}(\omega)|^2 /S(\omega )}] d\omega /2\pi$. The signal
$I_{sig} (\omega )=\int_0^\infty [I(t)-I (t\rightarrow\infty)]\exp
(i \omega t)dt$ corresponds to a deviation of the detector current
from its stationary value, and can be evaluated using Eqs.~(2), (3),
and~(5). We show in Fig.~3 how the integrated signal to noise ratio
behaves as a function of both $U$ and the ratio $\Gamma_R/\gamma$,
for $\Gamma_L=5 \Omega$ and $\gamma =\Omega$. A peak is observed
uniformly throughout the range of $U$ at $\gamma\approx
0.4\Gamma_R$. The signal to noise ratio depends weakly on $U$ when
$U/\Omega\gtrsim 15$, allowing good operation even at low values of
$U$. Comparing the performances of the DD and the SET detectors, we
see that the maximal signal to noise ratios obtained are comparable.
However, the SET reaches these values only in the asymmetric limit
$\Gamma_R\gg\Gamma_L$ \cite{gb}, while the DD detector's signal to
noise ratio is shown to be optimized without such a restriction.
\begin{figure}
{\centering{\epsfig{figure=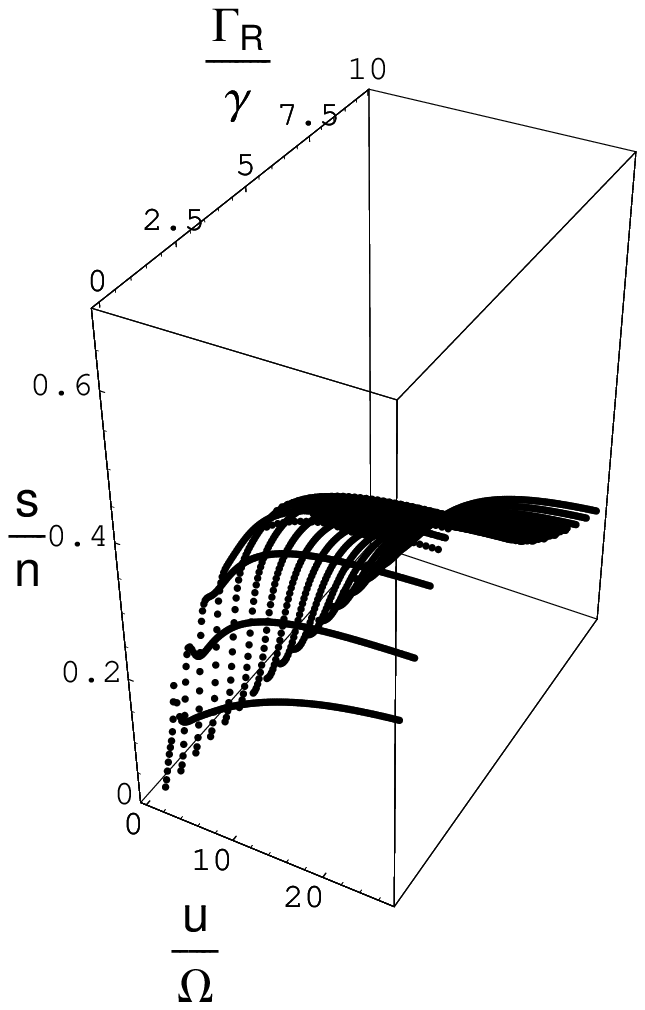,height=9cm,width=8cm,angle=0}}}
{\bf Fig.~3:} The integrated signal to noise ratio of the DD
detector. Here $\gamma=\Omega,\Gamma_L =5\Omega$.
\end{figure}

The results so far presented refer to a qubit that is positioned
near the right dot of the detector. One may ask what happens when
the qubit is positioned near the left dot. It was suggested in
Refs.~\cite{g4,gb} that the performance of any quantum detector
would improve if the detector were to operate mostly in the states
where there is no actual interaction with the qubit. By this
argument one can expect that putting the qubit near the left dot
would lead to poor performance. Indeed, in this case the
misalignment of the energy levels prevents electron propagation to
the right dot, so that the electron is pinned to the left dot. This
increases the weight of states where the detector interacts with the
qubit. On the other hand, if the qubit is located near the right dot
as in Fig.~1(B,B'), the same misalignment of levels localizes the
electron in the left dot, diminishing the occupation of the right
dot, Eq.~(\ref{a6}). As a result, the actual interaction
with the detector decreases and so it is expected to operate better.

Our conclusion about the asymmetry with respect to the qubit's
location can be confirmed by direct evaluation of the detector
efficiency, in the same way as presented in Figs.~2 and~3. We can
also confirm it in a different way by evaluating the power spectrum
of the detector current, $S(\omega )$, via Eq.~(\ref{a5}). This
displays a pronounced peak at $\omega = 2\Omega$, generated by the
qubit oscillations. It was argued in Ref.~\cite{kor} that the
peak-to-background ratio, $S(2\Omega )/S(\omega\to\infty )$, is a
measure of the detector efficiency. Figure~4 exhibits this ratio for
the two qubit positions as a function of $U$. As we increase the
misalignment of the levels we find that the two curves separate.
Thus, the setup with the qubit near the right dot is indeed more
effective, in accordance with our arguments. Note that the
peak-to-background ratio for this setup  depends weakly on $U$ for
$U/\Omega\gtrsim 15$. We have already noted that the signal-to-noise
ratio, which is another criterion of the detector efficiency, shows
a similar behavior.

\begin{figure}
{\centering{\epsfig{figure=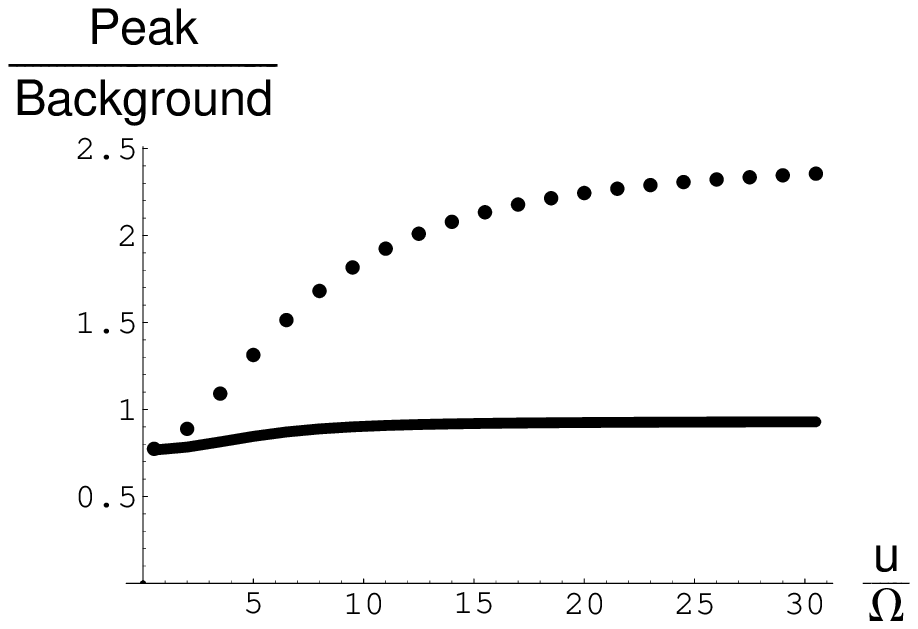,height=4cm,width=7cm,angle=0}}}\\
{\bf Fig.~4:} The peak to background ratio of the current power
spectrum  as a function $U$ for $\Gamma_L=5\, \Omega,\,
\Gamma_R=10\, \Omega$ and $\gamma=\Omega$. Solid line: qubit
positioned next to the detector's left dot, dotted: qubit positioned
next to the detector's right dot.
\end{figure}

We can show that the maximal value of the peak-to-background ratio
for the DD detector approaches 3 when $\Gamma_R,U\gg\gamma,\Omega$
(the dependence on $\Gamma_L$ is not essential). The same maximal
value was obtained for the SET detector\cite{gb}. Therefore, while
both detectors are sensitive measurement devices, they do not reach
the effectiveness of an ideal detector\cite{kor}.

In summary, we have proposed the use of a double dot structure for
the measurement of a charge qubit. We obtained a set of rate
equations describing the entire system and displayed the conditions
under which such a measurement is effective. We found the
measurement to be most sensitive when the detector operates mainly
in the states where no interaction with the qubit takes place. We
further demonstrated that, because of quantum interference effects
inside the detector, the stationary current is not determined solely
by the probabilities of the stationary qubit, but reflects the qubit
phase as well. 

\end{multicols}

\begin{references}
\bibitem{fn0} In fact, the use of a double-dot as a
sensitive charge detector has attracted considerable attention in
recent theoretical and experimental investigations\cite{tan,bre}.
\bibitem{fn1} In general the noise spectrum of the circuit current
is given by $\alpha S_L(\omega )+\beta S_R(\omega
)-\alpha\beta\omega^2 S_Q(\omega )$\cite{moz2}, where the
coefficients $\alpha,\beta$ with $\alpha +\beta =1$ depend on the
junction capacities, $S_{L,R}(\omega )$ are the noise spectrum in
the leads and $S_Q(\omega )$ is the charge correlation functions of
the DD. In this paper we take $\beta =1$ for the definiteness.
\bibitem{fn2} This condition for instance, is satisfied for an
``ideal'' point-contact detector\cite{g3,kor}. See also\cite{gb}.
\end{references}
\end{document}